\documentclass[prl, twocolumn, nofootinbib, floatfix]{revtex4-1}

\usepackage{amsmath}
\usepackage{graphicx}
\usepackage{dcolumn}
\usepackage{bm}
\usepackage{epsfig}
\usepackage{amssymb,latexsym,mathrsfs}
\usepackage{graphicx}
\usepackage{color}
\usepackage{hyperref}
\usepackage{xcolor}

\hypersetup{
    colorlinks=true,
    linkcolor=red,
    citecolor=blue,
}

\newcommand{\be}{\begin{equation}}
\newcommand{\ee}{\end{equation}}
\newcommand{\bs}{\begin{split}} 
\newcommand{\bea}{\begin{eqnarray}}
\newcommand{\eea}{\end{eqnarray}}

\newcommand{\Oe}{\Omega_{e}}
\newcommand{\lcdm}{$\Lambda$CDM} 
\newcommand{\neff}{N_{\rm eff}} 
\newcommand{\hfid}{H^2_{\rm fid}} 
\newcommand{\dl}{\delta}

\begin{document}

\title{New Constraints on the Early Expansion History} 
\author{Alireza Hojjati$^1$, Eric V.\ Linder$^{1,2}$, Johan Samsing$^{3}$} 
\affiliation{$^1$Institute for the Early Universe WCU, Ewha Womans 
University, Seoul 120-750, Korea\\ 
$^2$Berkeley Center for Cosmological Physics \& Berkeley Lab, 
University of California, Berkeley, CA 94720, USA\\ 
$^3$Dark Cosmology Centre, Niels Bohr Institute, University of Copenhagen, 
Juliane Maries Vej 30, 2100 Copenhagen, Denmark}

\begin{abstract}
Cosmic microwave background measurements have pushed to higher resolution, 
lower noise, and more sky coverage.  These data enable a unique test of the 
early universe's expansion rate and constituents such as effective number of 
relativistic degrees of freedom and dark energy.  Using the most recent 
data from Planck and WMAP9, we constrain the expansion history 
in a model independent manner from today back to redshift $z=10^5$.  
The Hubble parameter is mapped to a few percent precision, limiting early 
dark energy and extra relativistic degrees of freedom within a model 
independent 
approach to 2--16\% and 0.71 equivalent neutrino species respectively 
(95\% CL).  
Within dark radiation, barotropic aether, and Doran-Robbers models, the 
early dark energy constraints are 3.3\%, 1.9\%, 1.2\% respectively. 
\end{abstract}

\date{\today} 

\maketitle


Except for the last e-fold of cosmic expansion, our knowledge of the 
state of the universe arises directly only through measurements of the 
cosmic microwave background (CMB) radiation or indirectly (as in models 
of its influence on growth of large scale structure).  Recent CMB data 
\cite{planck15,wmap9} provides 
a clear window on an additional 10 e-folds of history (back to redshift 
$z=10^5$), a vast improvement in mapping the universe. 

The expansion rate, or Hubble parameter, is a fundamental characterization 
of our universe, and includes information on its matter and energy 
components, their evolution, and the overall curvature of spacetime.  
Moreover, the CMB encodes linear perturbations in the photons and the 
gravitational potentials they experience, providing sensitivity to the 
microphysics of components, e.g.\ their sound speed. 

These observations lead to constraints on quantities such as early dark 
energy and extra neutrino species or 
other relativistic degrees of freedom. 
However most analyses assume a specific model for these 
deviations, enabling stringent but model dependent constraints.  
In this Letter our 
approach is to map the cosmic state and history in as model independent 
fashion as practical, guided by the data.  We utilize the results of the 
principal component analysis of \cite{sls} to define localized bins of 
Hubble parameter in log scale factor that are most sensitive to the data, 
and then carry out a Markov Chain Monte Carlo (MCMC) analysis to constrain 
them.  
Finally we discuss the implications for 
dark energy, relativistic degrees of freedom, and spacetime curvature. 
For data we use the most recent CMB results from the Planck satellite 
\cite{planck15} and WMAP satellite \cite{wmap9}.

{\it Cosmic History Mapping --\/} 
For robust, model independent results we adopt a combination of 
principal component analysis (PCA) and binning.  This avoids assuming 
a specific functional form for the Hubble parameter or dark components 
and allows the data itself to inform where the greatest sensitivity lies. 
Such PCA on the Hubble parameter for projected mock CMB data was used 
in \cite{sls} to predict the strength of constraints at various epochs of 
cosmic history. 

This identification of the key epochs where physical conditions most 
affect the observations enables informed choice of bins in log scale 
factor to use in a MCMC fit.  Bins have several advantages over the 
raw PCA: 1) they are localized and can be clearly interpreted physically -- 
the Hubble parameter during a specific epoch, 2) they avoid negative 
oscillations that can cause unphysical results (while the sum of all 
PCs will give a positive, physical Hubble parameter squared, this is not 
guaranteed for a subset), and 3) they are well defined, not changing when 
new data is added.  

The Hubble parameter, or logarithmic derivative of the scale factor, 
$H=d\ln a/dt$, is then written as 
\be 
H^2(a)=\frac{8\pi G}{3}[\rho_m(a)+\rho_r(a)+\rho_\Lambda]\,[1+\delta(a)] \,, 
\ee 
where $\delta$ accounts for any variation from $\Lambda$CDM (cosmological 
constant plus cold dark matter plus standard radiation) expansion 
history, and $\rho_m$ is the matter density, $\rho_r$ the radiation density, 
and $\rho_\Lambda$ the cosmological constant density.  The bins in the 
deviation $\delta(a)$ are slightly smoothed for numerical tractability, 
with 
\be 
\delta=\sum_i \delta_i \left[\frac{1}{1+e^{(\ln a-\ln a_{i+1})/\tau}}-\frac{1}{1+e^{(\ln a-\ln a_i)/\tau}}\right] \,. \label{eq:delta} 
\ee 
Within bin $i$, $\delta=\delta_i$ and far from any bin $\delta=0$.  A 
smoothing length $\tau=0.08$ was adopted after numerical convergence tests. 
(A similar binned approach was used in \cite{linsmith,unique} to bound early 
cosmic acceleration.) 

We modify CAMB \cite{camb} to solve the Boltzmann equations for the photon 
perturbations in this cosmology.  The dark energy density contributed by the 
deviations $\delta$ and the cosmological constant term (which becomes 
negligible at high redshift) has an effective equation of state 
\be 
1+w=\frac{Q\delta}{1+\delta(1+Q)}\,(1+w_{bg}) 
-\frac{1}{3}\frac{1+Q}{1+\delta(1+Q)}\frac{d\delta}{d\ln a} \ , 
\label{eq:wddelta} 
\ee 
where $Q=(\rho_m+\rho_r)/\rho_\Lambda$ and $w_{bg}$ is the background 
equation of state of the combined matter and radiation (e.g.\ $1/3$ during 
radiation domination, transitioning to 0 
during matter domination).  Thus $w$ and $w'=dw/d\ln a$, entering into 
the Boltzmann equations, are defined fully by Eq.~(\ref{eq:delta}) for 
$\delta$.  We 
choose the associated sound speed to be the speed of light, as in 
quintessence dark energy, but explore variations of this later. 

Guided by the PCA of \cite{sls}, where the first few PCs show greatest 
sensitivity in $\log a\in[-4,-2.8]$, we choose bins $\dl_{1-5}$ in the 
logarithmic scale 
factor $\log a=[-5,-4]$, $[-4,-3.6]$, $[-3.6,-3.2]$, $[-3.2,-2.8]$, 
$[-2.8,0]$ so the finest binning is near CMB recombination at 
$a\approx10^{-3}$.  (Future CMB data could change the PCs, but we could 
keep the same bins, or not.)  The cosmological parameters we fit for are 
the six standard ones: physical baryon density $\Omega_b h^2$, physical 
cold dark matter density $\Omega_c h^2$, acoustic peak angular scale 
$\theta$, primordial scalar perturbation index $n_s$, primordial scalar 
amplitude $\ln(10^{10}A_s)$, and optical depth $\tau$, plus the five 
new deviation parameters $\dl_{1-5}$.  
Additional astrophysical parameters enter from the data, as discussed 
next.

{\it Constraints --\/} 
To constrain the cosmology with the data we use MCMC analysis, 
modifying CosmoMC \cite{cosmomc}.  The likelihood involves the 
temperature power spectrum from the two satellite experiments, and the 
E-mode polarization and TE cross spectrum from WMAP (the first Planck 
likelihood release does not include polarization, or the high multipole 
likelihoods from Atacama Cosmology Telescope \cite{act} or South Pole 
Telescope \cite{spt}; in the future such data should become available). 
Astrophysical nuisance parameters characterizing foregrounds (see 
\cite{planck15}) are marginalized over. 

\begin{figure}[htbp!]
\includegraphics[width=\columnwidth]{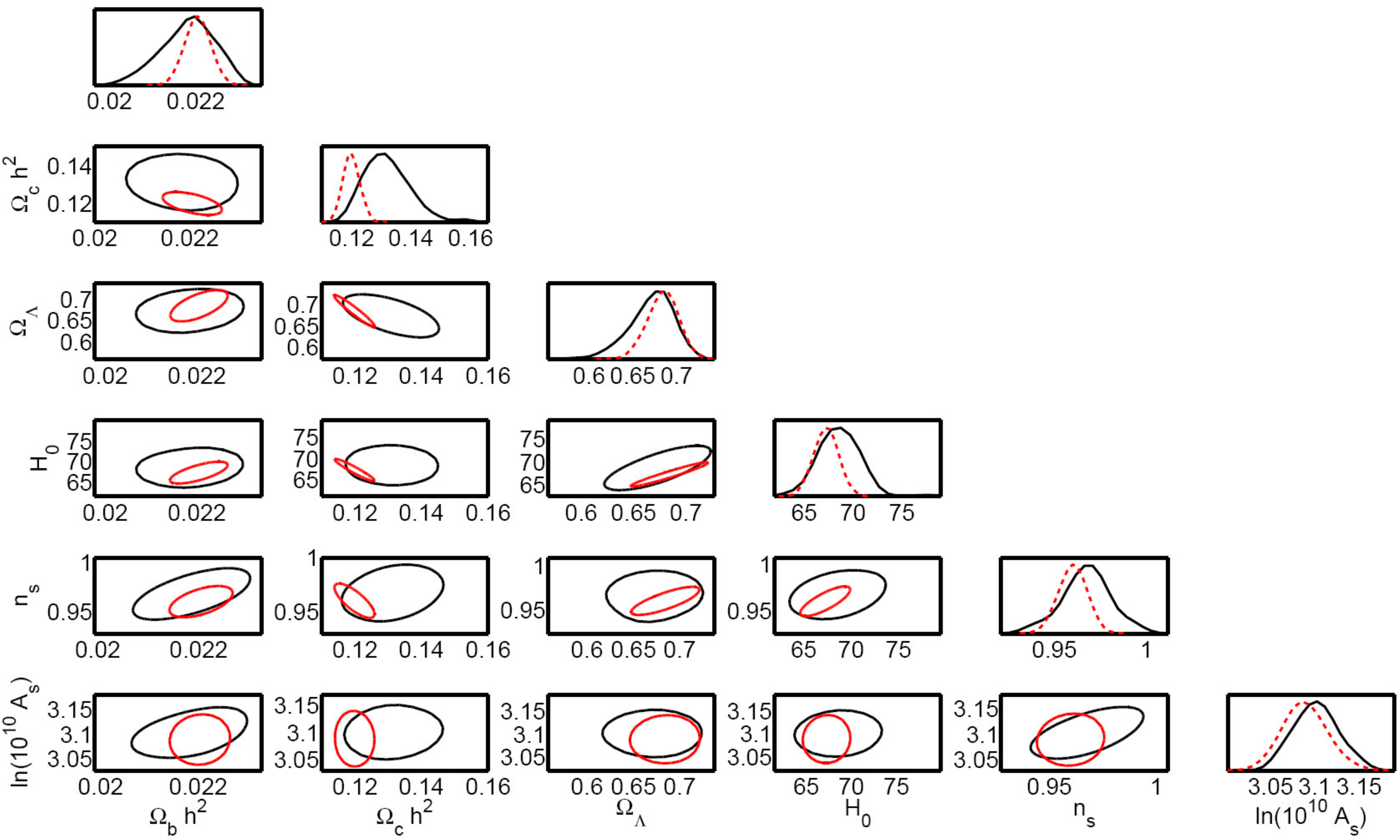}
\caption{Joint 68\% confidence contours on the standard cosmological 
parameters are shown when allowing for expansion history deviations from 
\lcdm\ (black), and fixing to \lcdm\ (smaller contours or dashed curves).  
Plots on the diagonal give the 1D marginalized probability distributions. 
}
\label{fig:std}
\end{figure}

Figure~\ref{fig:std} shows the constraints on the standard cosmological 
parameters, in the \lcdm\ case (fixing $\delta_i=0$) and 
when allowing variations in the expansion history (fitting for the 
$\delta_i$).  Here the Hubble constant $H_0$ replaces the $\theta$ 
parameter and we omit showing $\tau$.  Including the fitting for 
expansion history deviations induces roughly a factor of two larger 
marginalized 
estimation uncertainties for most of the standard cosmology parameters, 
and significantly shifts the cold dark matter density value.  This 
is due to the deviations in the Hubble parameter having similar effects on 
the expansion near recombination to those in matter, so $\delta$ takes the 
place of some of $\rho_m$.   We discuss this degeneracy further later. 
The best fit for the \lcdm\ case remains within the 68\% confidence contour 
when allowing expansion deviations.

\begin{figure}[htbp!]
\includegraphics[width=1\columnwidth]{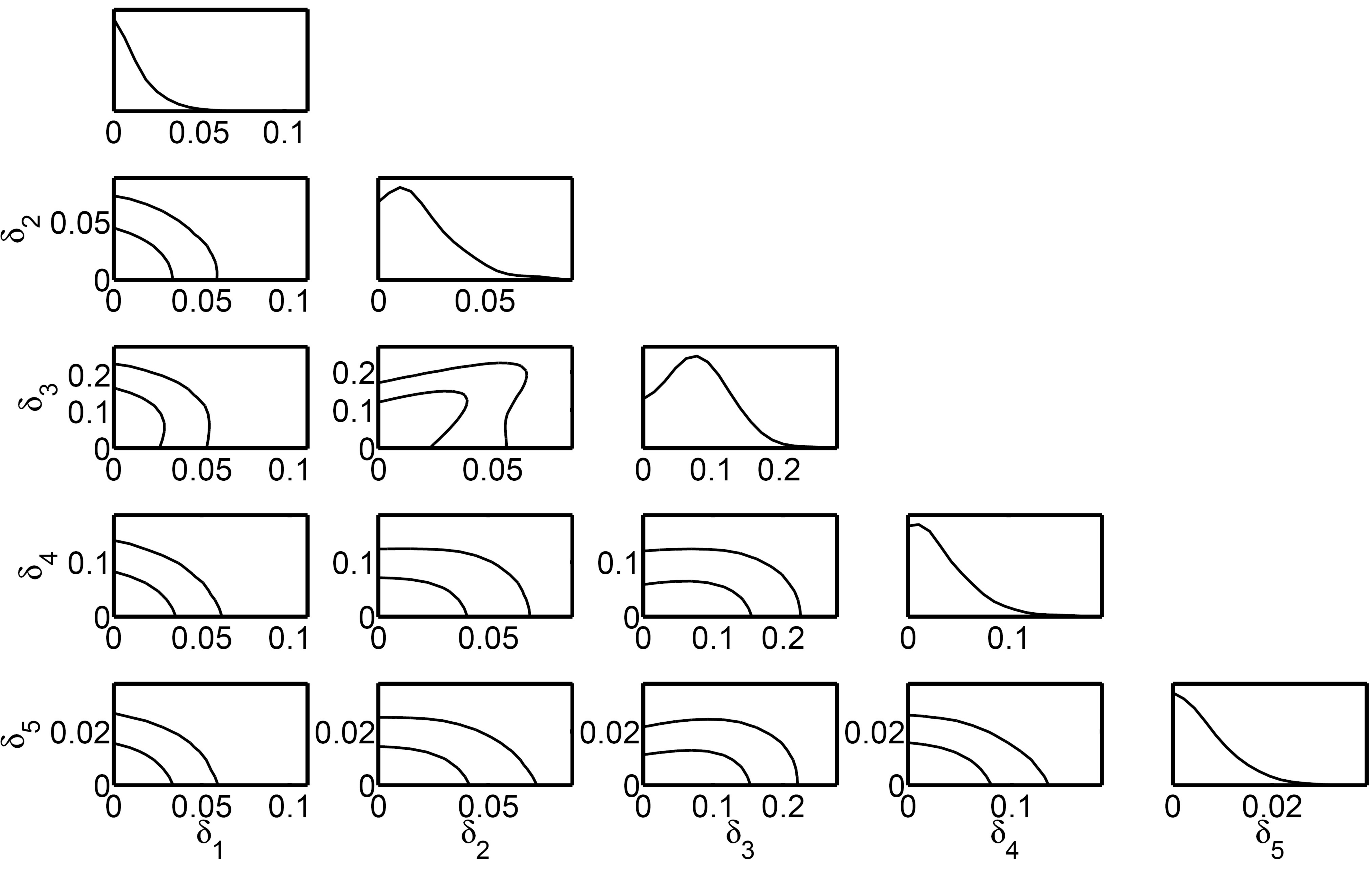}
\caption{Joint 68\% and 95\% confidence contours on the expansion deviation 
parameters are shown.  Plots on the diagonal give the 1D marginalized 
probability distributions. 
}
\label{fig:deltas}
\end{figure}

Figure~\ref{fig:deltas} shows the constraints on the expansion history 
deviations.  Note that to ensure positive energy density (and Hubble parameter 
squared) we restrict $\delta\ge0$, i.e.\ equal or more early energy density 
than in 
the \lcdm\ case (which has $\Omega_\Lambda\approx 10^{-9}$ at $a=10^{-3}$; 
allowing the limiting non-negative energy density $\delta\approx -10^{-9}$ 
would have negligible impact on the distributions). 
Note that these binned deviations do not have appreciable covariances 
with each other, with the correlation 
coefficients under 0.26 except for $\dl_2$--$\dl_3$ at 0.49.  
This is a useful feature adding near independence to 
localization, making their interpretation transparent, and is a result of the 
careful choice of bins based on the PCA of \cite{sls}.

Table~\ref{tab:delta95} gives the 95\% confidence upper limits on each 
expansion deviation parameter, showing that recent CMB data provides 
2--16\% 
constraints on the expansion history back to $z=10^5$.  The earliest 
bin, of $\delta_1$, is reasonably constrained despite being well before 
recombination, and should improve further when adding high resolution 
(high multipole $l$) 
measurements.  The second bin has equivalent constraints when taking into 
account its narrowness.  
Around recombination, however, $\delta_3$ and $\dl_4$ have 
looser bounds because all the standard cosmological parameters also 
enter strongly at this epoch, and so the increased covariance dilutes 
their estimation.  They have the two highest correlation coefficients, 
of 0.89 between $\delta_3$ and $\Omega_c h^2$ and $-0.76$ between $\dl_4$ 
and $\Omega_b h^2$.  Finally, the late, broad bin of $\delta_5$ has strong 
constraints.  These behaviors are all consistent with the pre-Planck, 
Fisher matrix predictions of \cite{sls} (see their Fig.~4).  Adding late 
time data or priors (which we avoid; see concluding section) 
can shrink some uncertainties by up to 60\%. 

The expansion history does not completely define the system of 
Boltzmann equations: the 
effective dark component can have internal degrees of freedom such as 
sound speed $c_s$ that determine the behavior of its perturbations and 
hence the gravitational clustering of the photons \cite{hu98}.  
Therefore we also show in Table~\ref{tab:delta95} 
the constraints when this sound speed is equal to that of a relativistic 
species ($c_s^2=1/3$), or is much smaller than the speed of light, 
cold dark energy with $c_s=0$.  The $c_s=0$ case has looser 
bounds, due to the additional influence on the photon clustering with the 
strengthened gravitational potentials, and covariance with matter parameters 
during matter domination.  
For the $c_s^2=1/3$ case, where the extra 
expansion rate corresponds to extra relativistic degrees of freedom, 
the constraints are weaker during radiation domination.  
This is a combination of the expansion deviation 
acting just like the photons, and a slight preference of the data for 
additional radiation energy density, in accord with previous hints that 
the number of effective neutrino species, 
$\neff$, might be greater than the standard model value of 3.046.  
Indeed, 
the mean value of $\delta_2=0.026$ in this 
case corresponds to $\Delta\neff=0.31$, in good 
agreement with the Planck values of $\neff=3.39$.  
Recall that $\Delta\neff$ denotes the equivalent number of relativistic 
neutrino species corresponding to the extra energy density.  Since $\delta_2$ 
is not in the fully radiation dominated era, we must translate it to the 
constant early dark energy density using Eq.~(25) of \cite{sls} and then to 
the asymptotic relativistic $\Delta\neff$ using Eq.~(6) of \cite{calabrese}. 

In all other parts of the article we keep $c_s=1$.

\begin{table}[!htb]
\begin{tabular}{l|ccccc} 
Case& $\delta_1(10^{-4.5})$&$\delta_2(10^{-3.8})$&$\dl_3(10^{-3.4})$&$\dl_4(10^{-3.0})$&$\dl_5(10^{-1.4})$\\ 
\hline 
$c_s^2=1$ & 0.036 & 0.050 & 0.160 & 0.095 & 0.018 \\
$c_s^2=1/3$& 0.053 & 0.054 & 0.067 & 0.038 & 0.013 \\
$c_s^2=0$  & 0.060 & 0.069 & 0.109 & 0.184 & 0.223\\
\end{tabular}
\caption{95\% confidence upper bounds are given for the expansion history 
deviations $\dl$, listed by the bin number and midpoint of the $\log a$ 
bins, for cases with different sound speeds. 
}
\label{tab:delta95} 
\end{table}

\begin{figure}[htbp!]
\includegraphics[width=\columnwidth]{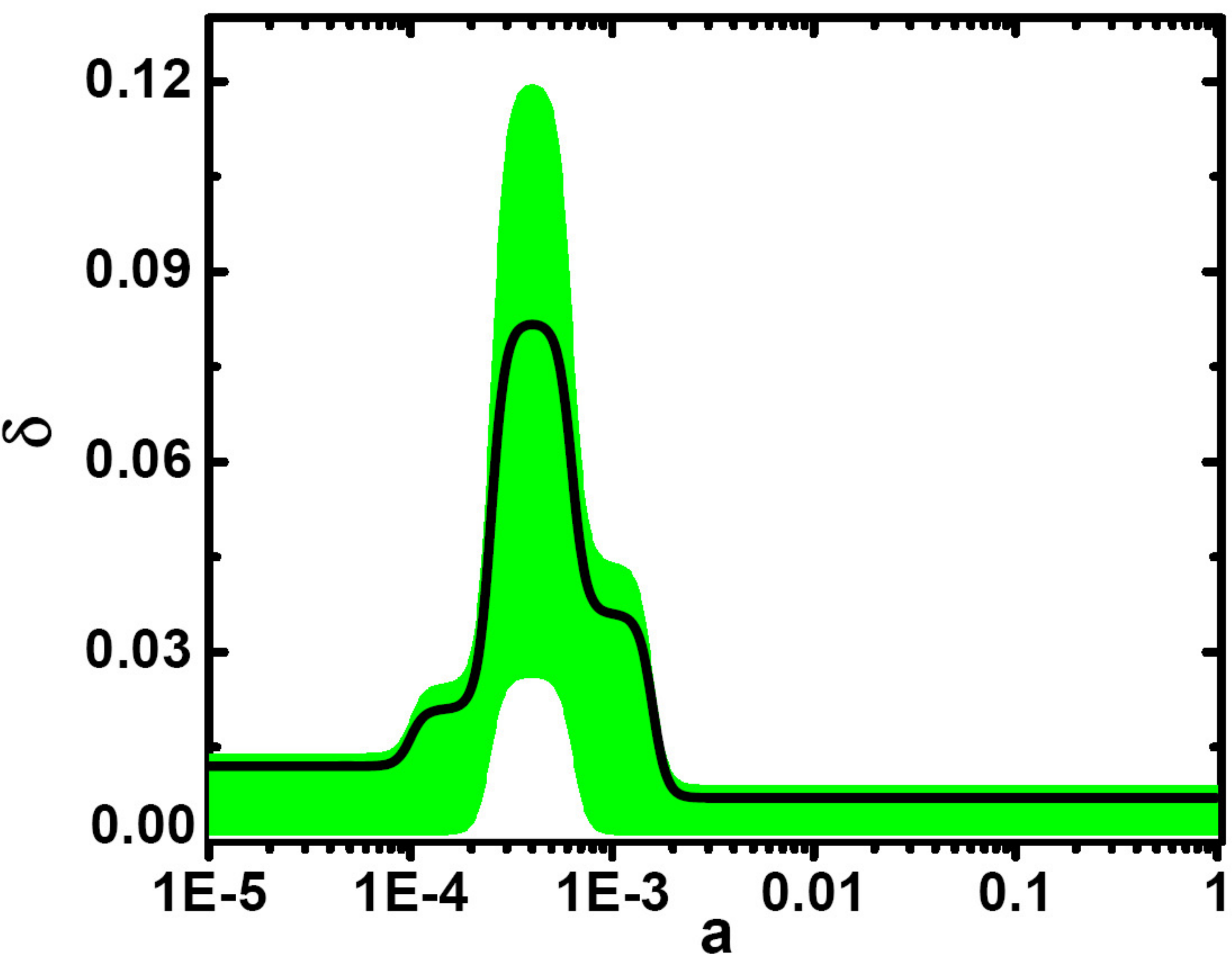}
\caption{Reconstruction of the expansion history deviations 
$\dl(a)$ from \lcdm\ is shown, with the mean value (solid line) and 68\% 
uncertainty band (shaded area).  
}
\label{fig:recon}
\end{figure}

Figure~\ref{fig:recon} shows the mean value and 68\% uncertainty band 
of the expansion deviations $\delta(a)$ given by the Monte Carlo 
reconstruction using the recent CMB data.  This figure represents the 
best current model-independent knowledge of the early expansion history 
of our Universe.  
Setting all $\dl_i=0$, i.e.\ \lcdm, is consistent with these results at 
the 95\% confidence level.  
The mean value does show a very slight preference for a faster expansion rate, 
as in early dark energy or extra relativistic degrees of freedom, 
before recombination.

{\it Physical Implications --\/} 
This analysis has been model independent, allowing individual epochs 
to float freely without assuming a functional form.  If we do assume a 
specific model, then constraints will in general be tighter, with each 
epoch having leverage on others through the restricted form.  

Three distinct families of early dark energy might be considered: 
where the early dark energy density rises, falls, or stays constant across 
CMB recombination.  These were investigated in \cite{sls} in terms of the 
(somewhat motivated) models of 
barotropic aether, dark radiation, and Doran-Robbers \cite{dorrob} 
forms, respectively (see \cite{sls} for more detailed discussion).  
We compute the constraints on the fraction $\Omega_e$ of critical density 
contributed by early dark energy 
(approximately equivalent to $\dl$) within each of these models (not using 
the $\dl_i$ bins), giving the results in Table~\ref{tab:3models}.  
(Note that Planck finds $\Omega_e<0.009$ at 95\% CL for the 
Doran-Robbers model when also including high multipole data \cite{planck16}.)

\begin{table}[!htb]
\begin{tabular}{c|ccc} 
& \ Aether \ & \ Dark Radiation \ & \ Doran-Robbers \ \\ 
\hline 
$\Omega_e$ & 0.019 & 0.033 & 0.012 \\ 
\end{tabular} 
\caption{The 95\% confidence level uncertainties are presented for 
three early dark energy models.  For small values, $\Omega_e\approx\dl$.  
The Doran-Robbers model has an additional parameter $w_0$; we find 
$w_0= -1.49^{+0.69}_{-0.57}$ (95\% CL). 
} 
\label{tab:3models} 
\end{table}

Two aspects of the models impact their detectability: the presence of 
the expansion history deviation at a sensitive epoch and its persistence 
over time, and its clustering behavior.  The common Doran-Robbers form has 
the tightest bounds (despite the extra parameter), due to its persistence 
pre- and post-recombination and its distinction from matter clustering 
since it has $c_s^2=1$.  The aether model only begins to deviate 
around recombination, and has $c_s^2=0$ so there is 
more covariance with the dark matter component.  Dark radiation has 
influence only before recombination and its $c_s^2=1/3$ makes it more 
covariant with the photons (and neutrinos).  A key conclusion is 
that early dark energy could in fact be more prevalent than apparent from 
bounds in the literature on the Doran-Robbers model.  

Since dark radiation density at early times scales like radiation, it acts 
like the addition of relativistic degrees of freedom.  Taking into account 
the definition of extra degrees in terms of the number of effective neutrino 
species $\neff$, the constraint on $\Omega_e$ within the dark radiation model 
translates to \cite{calabrese} 
\be 
\Delta\neff(a\ll a_{\rm eq})= 7.44\,\Oe/(1-\Oe) \ . 
\ee 
Thus $\Omega_e<0.033$ for the dark radiation model becomes $\Delta\neff<0.25$ 
at 95\% CL.  
This puts a tighter global bound on $\Delta\neff$ compared to our model 
independent value from $\delta_2$ before recombination ($\Delta\neff<0.71$ 
at 95\% CL, where again we have to account for $\delta_2$ not being in the 
fully radiation dominated era). 

Another implication of the expansion history is its relation to the 
spacetime itself.  The Ricci scalar curvature is the central quantity 
in the Einstein-Hilbert action for general relativity, and plays a key 
role as well in extensions to gravity such $f(R)$ theories.  The 
curvature history of the Universe has been explored from a theoretical 
perspective recently by \cite{caldwell}.  Since 
\bea 
R&=&3H^2\left[1-3w_{bg}\frac{\hfid}{H^2}-3w\frac{\dl H^2}{H^2}\right]\\ 
&=&3\hfid\left[1-3w_{bg}+\dl(1-3w)\right] \ , 
\eea 
observational constraints on $\dl$ (and hence $w$ through 
Eq.~\ref{eq:wddelta}) can be used to cast light on the curvature history.

{\it Conclusions --\/} 
We have used the recent advances in CMB data to constrain the fundamental 
quantity of the expansion history of our Universe.  The results from the 
model independent analysis bound deviations from \lcdm\ at 2--16\% 
(95\% CL), depending on the epoch.  
This constrains any deviations, whether due to, e.g., some form of dark 
energy or a nonstandard number of relativistic degrees of freedom.  It 
also relates directly to the Ricci spacetime curvature. 

Adding late time data that helps to constrain $H_0$ or $\Omega_m$, say, 
would help break the degeneracy around recombination that led to the loosest, 
16\% upper bound on deviations.  However, proper treatment of this would 
require many low redshift bins to reflect the density of the data, 
while our focus here is on the early expansion history. 

We regard the model independence of the analysis as a signal virtue; 
however we can also compare the bounds for specific early dark energy 
models.  For the barotropic aether, dark radiation, and Doran-Robbers models 
we derive 95\% CL limits of less than 0.019, 0.033, 0.012 in early dark 
energy density 
$\Oe$, respectively.  We emphasize that bounds appear tightest when assuming 
the conventional Doran-Robbers form, and so early dark energy should be 
not be thought ruled out based purely on constraining this model.  In terms 
of extra effective neutrino species the model independent 
results imply $\Delta\neff<0.71$ 
at 95\% CL.  

Future CMB data, such as the release of polarization data from Planck, 
ACTpol \cite{actpol}, PolarBear \cite{polarbear}, SPTpol \cite{sptpol} 
experiments will enhance our knowledge of the 
history back to $z\approx 10^5$.  Exploring the even earlier universe will 
require neutrino, dark matter, or gravitational wave astronomy.  
Late time probes will continue to map the last e-fold of cosmic expansion 
in greater detail. 
Over just a few years cosmological observations have taken us from order 
unity uncertainty (with $\gtrsim10\%$ in narrow epochs around recombination 
and today) to a few percent level knowledge over more than 10 e-folds of 
cosmic history.

{\it Acknowledgments --\/} 
We thank Stephen Appleby, Scott Daniel, and Tristan Smith for helpful 
discussions.  
AH acknowledges the Berkeley Center for Cosmological Physics for hospitality. 
This work has been supported by WCU Korea grant 
R32-2009-000-10130-0 
and the Director, 
Office of Science, Office of High Energy Physics, of the U.S.\ Department 
of Energy under Contract No.\ DE-AC02-05CH11231. 
The Dark Cosmology Centre is funded by the Danish National Research 
Foundation.


\end{document}